\documentstyle[preprint,aps,eqsecnum,epsf]{revtex}
\begin{document}
\preprint{\baselineskip 18pt{\vbox{\hbox{SU-4240-648(revised)}
\hbox{hep-th/9712247} \hbox{April,1997}}}}
\title{Anomalous Defects and Their Quantized Transverse
Conductivities}
\vspace{15mm}
\author{A.P. Balachandran, Varghese John,  Arshad Momen }
\vspace{15mm}
\address{ Department of Physics, Syracuse University,\\
Syracuse, NY 13244-1130, U.S.A.}
\vspace{1mm}
\author{Fernando Moraes}
\address{  Departamento de Fisica,\\
 Universidade Federal de Pernambuco,\\
50670-901 Recife, PE, Brazil.}
\maketitle
\begin{abstract}
Using a description of defects in solids in terms of three-dimensional
gravity, we study the propagation of electrons in the background of 
disclinations and screw dislocations.  We study the situations
where  there are bound states
that are effectively localized on the defect and hence can 
be described in terms of an effective $1+1$ dimensional 
field theory for the low energy excitations. In the case of screw dislocations,
we find that these excitations are chiral and can be  
described by an effective field theory of chiral fermions.
Fermions of both chirality occur even for a given direction of the 
magnetic field.  The ``net'' 
chirality 
of the system however is not always the same for a 
given direction of the magnetic field, but changes from one 
sign of the chirality through zero to the other sign 
as the Fermi momentum or  the magnitude of the magnetic flux is varied.
On coupling to an external electromagnetic field, the latter becomes
anomalous, and predicts novel conduction properties for these 
materials.
\end{abstract}
\vspace{5mm}

\newcommand{\be}{\begin{equation}}
\newcommand{\ee}{\end{equation}}
\newcommand{\bea}{\begin{eqnarray}}
\newcommand{\eea}{\end{eqnarray}}
\newcommand{\real}{{\rm l}\! {\rm R}}
\newcommand{\ra}{\rightarrow}
\newcommand{\tr}{{\rm tr}\;}
\newcommand{\al}{\alpha}
\newcommand{\del}{\Delta}
\newcommand{\pt}{\partial}
\newcommand{\Th}{\Theta}
\newcommand{\td}{\tilde{\del}}
\newcommand{\g}{\gamma}
\newcommand{\bt}{\beta}
\newcommand{\bchi}{\bbox{\chi}}
\section{Introduction}

The effect of defects on properties of solids has been a very active 
field of research\cite{nabarro}. 
The fact that the theory of defects in solids can be 
reformulated as a version of three-dimensional gravity has been discussed
in \cite{katanaev}. This formulation corresponds to looking at the 
continuum limit of a crystalline solid in which the static defect 
configuration is characterized by a non-trivial metric corresponding to a 
static spacetime.
There is a one-to-one correspondence 
between the classification of defects in terms of  Burgers vectors and 
different metric configurations. This formulation has the merit of 
highlighting 
the geometrical properties of defect configurations.

The properties of electrons interacting with these static 
defects
 can be studied by looking at  
their propagation in the background of these 
defects \cite{fernando}. 
In this paper, we will focus on situations where one can have bound states 
of electrons localized on line defects like the disclination and the screw 
dislocation. In this context, it is useful to recall 
that there is a very close similarity of this system
to that of particles in the 
background of cosmic strings and domain walls \cite{Witten,vilenkin}. 
It is known that there exist states localized on strings 
and domain walls, leading to interesting phenomena like gauge and 
gravitational anomalies.
The properties of states localized on domain walls in the context of 
condensed matter systems have also been studied in \cite{dagotto,kose}.

As mentioned above, we here study situations where there are 
bound states localized on line defects. 
In cases where one has chiral 
bound states, the effective description of low lying 
excitations at the defects
is given in terms of a $(1+1)$ dimensional massless chiral Dirac 
fermion theory.
Using well-known results about the gauge theory of
 chiral fermions in $(1+1)$
dimensions, one can 
show that there is a $U(1)$ anomaly\cite{schwinger}, 
leading to charge nonconservation in the 
$(1+1)$ dimensional effective action. One can also show that the charge 
nonconservation is exactly cancelled by the neglected 
bulk  modes\cite{Harvey}.
These modes contribute to the bulk current.
The latter leads to a net inflow of charge to the defect that 
precisely accounts for the charge violation given by the $U(1)$
anomaly. This effect of the bulk modes can be summarized in terms of 
effective Chern-Simons actions on  planes having the defect as 
boundary.

In order to give a simple but explicit example of the above scenario, 
we first study the
 Schr\"odinger equation of the electron in an external electromagnetic field
in the absence of any defect.
The coupling of the latter to the magnetic moment of the electron is also taken
into account. We assume that the gyromagnetic ratio of the electron is two. 
In this case, it is known that the
Dirac Hamiltonian in two dimensions can be squared to obtain the
Schr\"odinger equation with magnetic moment interaction. 
The corresponding  two dimensional 
(2d) Schr\"odinger Hamiltonian is the part of the 3d Schr\"odinger
Hamiltonian transverse to the defect. We first analyze the 2d
Dirac Hamiltonian
and its domains of 
self-adjointness. The latter  are found to be compatible with   
the domains for the
2d Schr\"odinger Hamiltonian when the derivatives of the Dirac 
wave function are in the same domain as the wave function itself.

In the absence of defects, 
the above 2d Dirac equation 
and the corresponding 2d
Schr\"odinger equation are known to have 
bound states of zero energy. The number of bound states is given by an 
index theorem and is related to the integer part of the flux through 
the plane. This is the Aharonov-Casher theorem \cite{AC}.
These bound states are localized in the region of the magnetic field.

We extend the Aharonov-Casher analysis to situations where there 
are special types of defects,
namely the edge disclination and the screw dislocation. We  find that 
similar results exist in this situation, that is  we can extend the counting 
of bound states to the situation where there is a conical deficit angle
and a screw defect.

Next we obtain a $(1+1)$ dimensional field
theory for the bound states as they move along the defect by linearizing 
the excitations about the 
Fermi surface. It is 
a $(1+1)$-dimensional chiral fermion theory for
the low energy excitations.
When this effective field theory is coupled to an 
external electromagnetic field,
there is then an anomaly \cite{schwinger}. 
We then proceed to look at the response of the system to a constant 
electric field applied in the direction of the defect 
and find  that there is a net inflow of charge from the planes 
transverse to the defect, in other words a 
``transverse Hall effect'', because of the anomaly.

The paper is arranged as follows. In section 2, we discuss the 
problem of the  motion of an electron on a plane (with or 
without a point 
removed ) in an external
 magnetic field and find the Dirac Hamiltonian. In section 3,
we do the partial-wave analysis for this Hamiltonian when the 
plane has a hole and find 
its domains of self-adjointness. We also find the zero-energy 
solutions for the Dirac Hamiltonian. We introduce the line defects,
namely disclinations and screw defects, in section 4.  We also
find the zero energy solutions in these background geometries, which 
are very similar to their flat space counterparts. These 
states turn out to be individually chiral in terms of their motion 
around the defect,
specifically, the screw defect. However, the total  
chirality 
of the system however is not always the same for a 
given direction of the magnetic field, but changes from one sign of 
the chirality through zero to the other sign 
as the Fermi momentum or the magnitude of the magnetic flux is varied.
In section 5, we write the (1+1)-dimensional
effective field theory for these chiral states and couple it to an external 
electric field  along the defect. This theory is anomalous as 
discussed above. This anomaly allows one to find the effective 
field theory in the bulk of the medium where the defect is embedded. It 
is given by 
Chern-Simons terms defined on the half-planes which have the line defect 
as the boundary. Finally, in
the concluding section 6,  we discuss other physical effects 
that can be deduced 
using this reformulation of defects in terms of non-trivial 
metric solutions of the gravitational equations.

\section{ The Hamiltonian on the plane}
In this section we will be discussing the 
motion of an electron in an external magnetic field  along the 
$z$-direction. We will consider the 
Schr\"odinger equation for a spin half 
particle with a magnetic moment interaction in this ambient space.

 The line element describing the Euclidean geometry of $\real^3$ 
 is given by
\be
ds^2= ( dr^2 + r^2 d \theta^2) + dz^2,
\label{1.1}
\ee
$r, \theta, z$ being the cylindrical coordinates. 
We will see later that various  defect geometries 
can also be described by  
line elements like the above one.

The spatial Laplacian is
\be
\del = [ \frac{1}{r} \frac{\pt}{\pt r}( r \frac{\pt}{\pt r} ) 
+ \frac {1}
{ r^2}\frac{\pt^2}{\pt \theta^2} + \frac{\pt^2}{\pt z^2} ]~ .
\label{1.4}
\ee

We now split $\del$ according to  
\be
\del = \del_T + \del_Z = \del_T + \frac{\pt^2}{\pt z^2}.
\label{1.5}
\ee

In the
presence of a magnetic field along the $z$-direction and also a magnetic 
moment interaction, the electronic  wavefunction on the 
plane satisfies the Pauli equation
\be
(-\frac{1}{2m^*}\bar{\del}_T - \frac{\mu}{2} {\sigma \cdot B}) \Psi = E \Psi
\label{1.7}
\ee
where $\bar{\del}_T = D_i D_i$
 is the transverse Laplacian constructed from the 
gauge-covariant derivative
\be
D_i \equiv \pt_i - i A_i ~,
\label{1.7a}
\ee
and $\mu $ and $m^*$ are the electronic 
magnetic moment and effective mass 
respectively. Notice that we have chosen our units by requiring 
the electronic charge $e$ to be 1. We have also chosen a gauge where 
$A_0= 0$.

When $\mu = \frac{1}{m^*}$, the Hamiltonian in (\ref{1.7})
 can be obtained by 
squaring the two-dimensional massless Dirac Hamiltonian. We will be dealing 
with this Dirac Hamiltonian below.

For regular Cartesian coordinates our choice of the $\gamma$ matrices 
are 
\be
\gamma^t \equiv \gamma^0 = \sigma_3; \qquad \gamma^x = -i \sigma_2; \qquad
\gamma^y = i \sigma_1,
\label{new0}
\ee
where $\sigma_i$ are the standard Pauli matrices. From here one can readily 
find the $\gamma$ matrices for the polar basis :
\be
\gamma^0 = \sigma_3, \qquad 
\gamma^1 =(\bbox{\gamma \cdot \hat{r}})
=  \left(\begin{array}{cc} 0 &-e^{-i\theta} \\
e^{i\theta}&0 \end{array} \right ), \qquad 
\gamma^2 = (\bbox{\gamma \cdot \hat{\theta}})
=\left(\begin{array}{cc} 0 & ie^{-i\theta} \\
ie^{i\theta} & 0 \end{array} \right ). 
\label{new1}
\ee 

We assume that $B$ is time-independent. We can then choose a gauge where 
both $A_0$ and $A_r$ are zero. In this gauge,
the transverse Dirac Hamiltonian is identified via
the equation $ i \frac{\partial}{\partial t} \Psi = H_D \Psi$ 
where
\bea 
H_D &=& -i \g^0( \g^x D_x + \g^y D_y) \nonumber \\\
&=& \g^0 [-i (\bbox{\g \cdot {\hat{r}}})  \partial_r  
- i\frac{1}{r} (\bbox{\g \cdot {\hat{\theta}}})(\partial_\theta - i r 
A_\theta) ] \nonumber 
\\
&=&  \left( \begin{array}{cc} 0 & i e^{-i\theta}
\left[ \frac{\pt}{\pt r} +\frac{1}{r}(-i \frac{\pt}
{\pt \theta} - r A_\theta ) \right] \\
 ie^{i\theta} \left[ \frac{\pt}{\pt r} -\frac{1}{r}(-i \frac{\pt}
{\pt \theta} -r  A_\theta ) \right] & 0 
\end{array}
\right).
\label{1.12}
\eea
This Hamiltonian is formally symmetric with respect to the 
measure $r ~dr~ d \theta$. 

Squaring this operator leads to the Hamiltonian (\ref{1.7}) with $\mu= 
\frac{1}{m^*}$ :
\be
H = \frac{1}{2 m^*} H_D H_D.
\label{1.13}
\ee

Next we proceed to study the boundary conditions at the defect at $r=0$
appropriate for this Dirac operator.

\section{The Dirac Equation on $\real^2 - \{0\}$ }

The massless Dirac equation on a plane with a point removed 
admits a one-parameter family of boundary conditions at $r=0$.
This can be seen as follows.

The Dirac operator $-i \bbox{\sigma \cdot \nabla}_T$
defined (in the absence of the magnetic field )
on the two-dimensional 
plane,
is self-adjoint on a domain $\cal{ D}$  if and only if \cite{simon} 
\bea
\int_0^\infty r dr \,\int^{2\pi}_0 d\theta 
\left[ \bchi^\dagger \{-i \bbox{\sigma \cdot \nabla}_T \Psi\} - 
\{-i \bbox{ \sigma \cdot \nabla}_T \bchi\}^\dagger \Psi \right ] = 0 \nonumber\\
\mbox {for}~~ \Psi \in \cal {D}
\Leftrightarrow \bchi \in \cal {D},
\label{2.4}
\eea
where
\be
\bbox{ \nabla}_T \equiv \bbox{\hat{r}} \frac{\partial}{\partial r} + 
\bbox{\hat{\theta}}\frac{1}{r}\frac{\partial }{\partial \theta}.
\label{tdi}
\ee
Using Stokes' theorem and assuming that the wavefunctions fall off 
appropriately
at spatial infinity, one ends up with the following 
equation at the boundary ( at $r=0$):
\be
\lim_{r \ra 0}  \int d \theta~~ r \bchi^\dagger{\cal  A}~ \Psi = 0 \qquad
\mbox {for}~~ \Psi \in \cal {D}
\Leftrightarrow \bchi \in \cal {D} .
 \label{2.5}
\ee
Here, 
\be
{\cal A} \equiv -i \bbox{ \sigma \cdot \hat{r}} = 
\left( \begin{array}{cc} 0 & -ie^{-i \theta} \\
-i e^{i \theta} & 0 \end{array} \right).
\label{2.6}
\ee
The eigenvalues of the operator ${\cal A}$ are $+ i$ and $-i$ and the 
corresponding eigenvectors normalized to $1$ are 
\be
\frac{1}{\sqrt{2}}\left( \begin{array}{c}1\\-e^{i \theta} \end{array}\right),
\qquad \mbox{and} \qquad 
\frac{1}{\sqrt{2}}\left( \begin{array}{c}1\\e^{i \theta} \end{array}\right)
\label{2.7}
\ee
respectively.
It follows that the boundary conditions at $r=0$ compatible with 
self-adjointness of $- i\bbox{\sigma \cdot \nabla}_T$
 are parametrized by $e^{iK} \in U(1)$ [ $K$ being real]  and 
are given by \\
\be
\lim_{r \ra 0}
\sqrt{r} \Psi(r,\theta) \equiv 
\left( \begin{array}{c} \psi_1 \\ \psi_2 \end{array} \right)
(0,\theta) = \alpha(\theta)
 \left[ \left( \begin{array}{c}1\\e^{i \theta} \end{array}\right) 
+ e^{iK}  
\left( \begin{array}{c}1\\-e^{i \theta} \end{array}\right)\right],
\label{2.8}
\ee
where $\alpha(\theta)$ is any smooth 
function on $S^1$. The parameter $K$ (mod $2 \pi$) parametrizes the various 
boundary conditions. One can readily see that in terms of  $K$,
\be
 \cos(\frac{K}{2})\psi_2(0, \theta)= 
-i e^{i \theta}\sin(\frac{K}{2}) \psi_1(0, \theta) .
\label{2.9}
\ee
When $K$ is $0$ or $\pi$, this shows that the components 
$\psi_2(0, \theta)$ or $\psi_1(0, \theta)$ vanish respectively and the 
solutions turn out to be ``chiral'' at the origin.

 It is important that the Laplacian remains self-adjoint under the chosen 
conditions. 
We must therefore augment (\ref{2.8}) by the 
same types of 
boundary condition on the radial derivatives too,
\be
\lim_{r \ra 0}
\sqrt{r} \partial_r \Psi(r,\theta) 
 = \beta(\theta)
 \left[ \left( \begin{array}{c}1\\e^{i \theta} \end{array}\right) 
+ e^{iK}  \left( \begin{array}{c}1\\-e^{i \theta} \end{array}\right)\right].
\label{2.8a}
\ee
Here $\beta(\theta)$ is any smooth function on $S^1$. [ It need not be 
$\alpha(\theta)$.] Equations (\ref{2.8}) and (\ref{2.8a}) lead to
\bea
\int r dr\, d\theta 
\left[ \bchi^{\dagger} {\del}_T \Psi - ({\del}_T 
\bchi)^{\dagger}\Psi\right]
= \lim_{r \ra 0} \int r~d \theta  [ (\chi^\dagger_1 \partial_r \psi_1 + 
\chi^\dagger_2 \partial_r \psi_2)
 - (\partial_r \chi^\dagger_1 \psi_1 + \partial_r 
\chi^\dagger_2 \psi_2)] = 0 \nonumber \\
\mbox {for}~~ \Psi \in \cal {D}
\Leftrightarrow \bchi \in \cal {D}. 
\label{2.9a}
\eea
Here ${\cal D} \equiv {\cal D}(e^{iK})$ is the 
set of functions fulfilling both the conditions (\ref{2.8}) and 
(\ref{2.8a}). Thus the Laplacian is also  self-adjoint with (\ref{2.8})
and (\ref{2.8a}).

We now include the magnetic field. It can easily be checked that the
self-adjointness conditions
discussed above remain unaffected as long as we are not dealing with 
singular magnetic field configurations.

We will assume that our magnetic field is non-vanishing only  
over a compact region enclosing the defect. For simplicity we will in 
fact assume that the magnetic field
is a  constant within a circle of radius $a$ and zero 
outside. 
( The magnetic field can be represented by a delta function in the limit
of $a=0$,
 this case has been treated in detail 
by Moroz \cite{moroz}.) Hence, our vector potential will be chosen to have 
the 
form
\bea
A &=& A_r dr + A_\theta r d \theta, \nonumber \\
A_r &=& 0, \qquad \qquad 
A_\theta = \frac{Br}{2}\tilde{\theta} ( a-r) + \frac{\Phi}{2\pi r} \tilde{
\theta} (r-a),\nonumber \\
dA &\equiv&B_z r dr ~d\theta= B \tilde{\theta}(a-r)d^2 x = B\tilde{\theta}(a-r) ~r dr 
d\theta 
\label{2.10}
\eea
where 
\be
\Phi= \pi a^2 B
\label{2.10a}
\ee
is the total flux and  $\tilde{\theta}$ the step function. [ We
denote it by $\tilde{\theta}$ to avoid duplication of notation.]

As the magnetic field is cylindrically symmetric,
the angular momentum for the Dirac particle is a good quantum
number. Accordingly, 
  the conditions (\ref{2.8}) and (\ref{2.8a}) 
can by analyzed in terms of the simultaneous 
eigenfunctions of $H_D$ and  the total
angular momentum operator  $J = L + \frac{1}{2} \sigma_3$
where $L=-i\frac{\pt}{\pt \theta}$ is the orbital angular momentum 
operator. The eigenfunction of the Dirac 
Hamiltonian  $H_D$ with total angular momentum $j$ 
 has the form
\be
\Psi_j (r, \theta ) = \left(\begin{array}{c}e^{i(j-\frac{1}{2})\theta} u_j(r)\\
e^{i(j+\frac{1}{2})\theta}v_j(r) \end{array}\right)\equiv
\left(\begin{array}{c}e^{im \theta} u_{m+\frac{1}{2}}(r)\\
e^{i(m+1)\theta}v_{m+\frac{1}{2}}(r) \end{array}\right)~,
 \label{2.1}
\ee
where $m = j-\frac{1}{2}$.
In terms of $u_{m+\frac{1}{2}}$ and $v_{m+\frac{1}{2}}$, the boundary condition
(\ref{2.9}) reads as
\be
\sqrt{r} \cos (\frac{K}{2})v_{m+\frac{1}{2}}(r)=
\sqrt{r} \sin (\frac{K}{2})u_{m+\frac{1}{2}}(r).
\label{new}
\ee

 The spectrum of angular momentum depends on the 
(quasi-)periodicity of the wave function under the $\theta \ra \theta + 2\pi$ 
translation,
\be
\Psi (r, \theta + 2\pi ) = e^{i \lambda } \Psi ( r, \theta)
\label{period}
\ee
which implies that 
\be
m \in \frac{\lambda}{2\pi } + n , \qquad n \in Z \!\!\!Z .
\label{spect}
\ee
The  Hamiltonian $H_D$, 
restricted to  
the $m$-th partial wave  $\Psi_m(r) \equiv 
\left(\begin{array}{c}u_{m+\frac{1}{2}} \\
v_{m+\frac{1}{2}} \end{array} \right)$ ,  acquires the form 
\bea
H_{D}^{(m)} &=& \left( \begin{array}{cc} 0 & i
\left[ \frac{\pt}{\pt r} +\frac{1}{r}(m+1- 
 r A_\theta ) \right] \\
 i
\left[ \frac{\pt}{\pt r}  -\frac{1}{r}(m
 -  rA_\theta ) \right] & 0 
\end{array}
\right) \nonumber \\
&\equiv& \left( \begin{array}{cc} 0 & D \\
D^\dagger &0\end{array} \right)~.
\label{2.2}
\eea
On squaring $H_D^{(m)}$, one obtains the Schr\"odinger operators 
\bea
DD^\dagger &=& - \bar{\del}_T^{(m)} - B_z, \nonumber \\
D^\dagger D &=& - \bar{\del}_T^{(m+1)} + B_z,  \nonumber \\
B_z &\equiv& \frac{1}{r}\partial_r ( r A_\theta), 
\label{2.3}
\eea
 $\del_T^{(m)}$ being the radial Laplacian including the gauge connection
(which is a function of $r$ only),
\be
\del_T^{(m)}= \frac{\partial^2}{\partial r^2} + \frac{1}{r} \frac{\partial}
{\partial r} - \frac{ ( m - r A_\theta )^2}{r^2}
\label{2.35}
\ee 
and $B_z$ is 
the component of the magnetic field in the third direction, 
that is, in the direction normal to the plane.

  We now  proceed to find zero energy solutions of the Dirac equation 
satisfying the boundary condition (\ref{2.8}). They automatically satisfy 
the boundary condition (\ref{2.8a}) and hence are zero energy solutions of the
Laplacian too.

For a generic $A_\theta$, the zero energy eigenfunctions of 
$H_D^{(m)}$ [ignoring the boundary condition (\ref{new}) and 
square integrability for the moment] are given by
\be
\Psi_m (r) = 
\left( \begin{array}{c} C r^{m} 
e^{- \int_0^r A_\theta(r')dr'}\\
D r^{-(m+1)} 
e^{ \int_0^r A_\theta(r')dr'} \end{array}
\right )
\label{2.11}
\ee

There is a bound on the value of $m$ 
from the requirement of square integrability of the eigenfunctions.
Let us assume that $\Phi$ is positive (i.e. the magnetic
field is pointing along positive $z$-axis). 
Using the asymptotic forms
\be
\int^r_0 A_\theta (r') dr' = \left\{ \begin{array}{l} 0 \qquad \mbox{as 
$r$ $\ra$ 0}, \\
\frac{\Phi}{2\pi}\ln r \mbox{ as  $r$ $\ra \infty$}
\end{array} \right.
\label{2.12}
\ee
which follow from (\ref{2.10}),
 the solutions (\ref{2.11}), the condition (\ref{new}) and the 
requirements of square-integrability of the wavefunctions at $r = 0$
and $r \ra \infty$, we obtain the wave function 
\be
\Psi_m(r)  = 
\left( \begin{array}{c}C r^{m} 
e^{- \int_0^r A_\theta(r')dr'}\\
0 \end{array}\right)
\label{psi1}
\ee
and  the following bounds on $m$ :
\bea
\mbox{a)  When } e^{iK} \neq  1 \quad:  \qquad -\frac{1}{2} 
< m < (\frac{\Phi}{2\pi} -  1 ), \nonumber \\
\mbox{b)  When } e^{iK} = 1 \quad : \qquad -\frac{1}{2} 
\leq m < (\frac{\Phi}{2 \pi} - 1 ). \nonumber \\
\label{2.13}
\eea
These bounds imply that 
\be
\Phi > \pi.
\label{2.13.5}
\ee
Thus, there are about $[ \frac{\Phi}{2\pi} - \frac{1}{2}]$ independent 
spin polarized zero energy solutions, $[\xi]$ denoting the largest integer 
not exceeding $\xi$. [ The precise number depends on the 
value of $e^{iK}$ and the allowed values of $m$.]  
It is important to note that the parameter $K$ determines whether 
the mode $m=-\frac{1}{2}$ is a bound state or not. Of course this value of
$m$ is allowed only if that $\lambda = \pi ( \mbox{mod} 2 \pi)$ as well.

On the other hand, when $\Phi $ is negative, we find the wave function 
\be
\Psi_m (r) = 
\left( \begin{array}{c}0\\
D r^{-(m+1)} 
e^{ \int_0^r A_\theta(r')dr'} \end{array} \right)
\label{psi2}
\ee
where $m$ is now bounded as follows: 
\bea
\mbox{c)  When } e^{iK} \neq   - 1 \quad :  \qquad -\frac{1}{2}
> m >\frac{\Phi}{2\pi}, \nonumber \\
\mbox{d)  When } e^{iK} =    - 1 \quad : \qquad   -\frac{1}{2}
\geq m >\frac{\Phi}{2\pi} . \nonumber \\
\label{2.13a}
\eea
As a consequence, 
\be
\Phi < -\pi.
\label{2.13a.5}
\ee

Note that no zero energy  bound states exist when $\Phi \leq |\pi|$.

We next proceed to extend this analysis to three dimensions and in 
particular to  dislocations and screw 
defects.

\section{Disclination and Screw Dislocation}

\subsection{ Disclination}

The above analysis can be extended very easily to the case when we have a 
conical defect or a screw dislocation in our background. In the
presence of a 
conical defect ( which one can create by simply cutting out a wedge from 
a plane and identifying [pasting together] the edges or by inserting 
a wedge),  the geometry of 
the defect can be represented  by the metric \cite{katanaev}
\be
ds^2= (dr^2 + \al^2 r^2  d\theta^2 ) + dz^2
\label{cone1}
\ee
where the coordinates $r,\theta,z$ are the standard cylindrical coordinates
[ with $\theta =0$ and $\theta = 2\pi$ being identified as usual ]
and $2\pi ( 1- \al)$ is the opening angle of the 
cutout wedge. Note that $0 \leq \al \leq 1$. 
In this coordinate system, the Laplacian can be written as
\be
 \del = \frac{1}{r} \frac{\pt}{\pt r}( r \frac{\pt}{\pt r} ) + \frac {1}
{\al^2 r^2}\frac{\pt^2}{\pt \theta^2} + \frac{\pt^2}{\pt z^2} ~.
\label{cone2}
\ee

It is convenient to rescale the coordinates as follows :
\bea
r \ra R =  r, \nonumber \\
\theta \ra \Th = \al \theta, \nonumber \\
z \ra Z =  z .
\label{cone3}
\eea

 The Laplacian then becomes 
\be
\td =  [ \frac{1}{R} \frac{\pt}{\pt R}( R \frac{\pt}{\pt R} ) 
+ \frac {1}
{ R^2}\frac{\pt^2}{\pt \Th^2} + \frac{\pt^2}{\pt Z^2}].
\label{cone4}
\ee

Although $\td$ looks like the flat space Laplacian, 
the range of $\Th$ is different,  $ 0 \leq \Th \leq 2\pi \al$. 
As a consequence, the (quasi)periodicity of 
the wavefunction now reads
\be  
\Psi ( R, \Th + 2\pi \al ) = e^{i \lambda}
\Psi ( R, \Th).  
\label{cone5}
\ee
We also demand the identical periodicity condition on the radial 
derivative to ensure that $\td $ is self-adjoint.

For the condition (\ref{cone5}), the spectrum of the ``angular momentum'' 
$-i\frac{\partial}{\partial \Theta}$ gets
 quantized in units
of $ \frac{1}{\al}(m+\frac{\lambda}{2\pi})$ , where $m
 \in Z\!\!\!Z^+$.
The method of counting the bound states is the 
same as in section 3, but for the change  $m \ra \frac{m}{\al}$. This 
incidentally also shows that the number of bound states can change in the presence of a 
disclination.

\subsection{ Screw Dislocation}

Next we move on to the screw dislocation.
The screw dislocation can be characterized by the metric 
\bea
ds^2 &=&( dr^2 +  \al^2 r^2 d\theta^2 )+(dz+\bt d\theta)^2 
\equiv g_{ij} d \xi^i 
d\xi^j, \nonumber \\
d \xi^1 &=& dr, \quad d \xi^2 = d \theta, \quad  d\xi^3 = dz
\label{s2}
\eea
where as usual $r,\theta,z$ are the cylindrical 
coordinates for $\real^3$.

The matrix $[g_{ij}]$ and its inverse $[g^{ij}]$ can be written as 
\be
\left[g_{ij}\right]= \left( \begin{array}{ccc} 1 &~~0& ~0 \\
0& (\al^2 r^2 + \bt^2)  & \bt \\
0 & \bt & 1 \end{array} \right), \qquad \qquad \left[g^{ij}\right] = 
\left( \begin{array}{ccc} 1 &~~~~0 &~~~~0 \\
0 &~~~~~ \frac{1}{\al^2 r^2} & ~~~~~\frac{-\bt}{\al^2 r^2} \\
0& ~~~~~~\frac{-\bt}{\al^2 r^2} &~~~~~ (1+ {\frac{\bt^2}{\al^2 r^2}})
\end{array}\right) 
\label{s2a}
\ee
respectively.

The Laplacian after a rescaling as in  (\ref{cone3}) is
\bea
\del_1 &=& \frac{1}{\sqrt{g}}\pt_i (g^{ij}\sqrt{g}\pt_j) \nonumber \\
&=& [ \frac{1}{R}(\frac{\pt}{\pt R}  R \frac{\pt}{\pt R} ) + \frac{1}{R^2}
( \frac{\pt}{\pt \Th} - \frac{\bt}{\al}\frac{\pt}{\pt Z})^2 
+ \frac{\pt^2}{\pt Z^2}],\\
\label{s3}
g &\equiv& det [ g_{ij}]. \nonumber 
\eea

Wave functions in the domain of this Laplacian fulfill (\ref{2.9}).

It may be emphasized here that in the presence of a  screw defect, the 
geodesics of  the metric $[g_{ij}]$ give the free propagation of 
low energy electrons.\footnote
{By making the coordinate change $Z \ra Z'=Z +\frac{\bt}{\al} 
\Th$, this problem can be 
mapped into that of a Laplacian on flat space, but with a non-trivial boundary 
condition. Specifically, the quasi-periodicity of the wavefunction under 
$\Th \ra \Th + 2\pi \al$ translates into
$$
\Psi ( R, \Th , Z') =e^{i \lambda}
 \Psi ( R, \Th + 2\pi \al , Z'+ 2\pi \bt). \nonumber $$ 
}

Next we add a magnetic field in the $Z$-direction. This is done as 
before, by replacing $\pt_{i}$ by $\pt_i -iA_i$
where $A_R=0; A_\Th = \frac{1}{\al}
[\tilde{\theta}(a-R)\frac{BR}{2}+ \frac{\Phi}{2\pi R}
\tilde{\theta}(R-a)]$ and also introducing a  magnetic moment 
interaction. Note that the potential has been suitably scaled 
so that the total flux is still $\Phi$. The Hamiltonian then reads
\be
H = -\frac{1}{2m^*}[\frac{1}{R}\frac{\pt}{\pt R}( R 
\frac{\pt}{\pt R} ) +\frac{1}{R^2}{\cal D}_\Th^2 
+\frac{\pt^2}{\pt Z^2}] - \frac{1}{2m^*}\bbox{\sigma\cdot B}
\label{def1}
\ee
where ${\cal D}_{\Th}= (\pt_{\Th} -iR A_{\Th}- \frac{\bt}{\al} 
\pt_{Z})$.

The zero energy eigenfunctions of $H$ are found as before, by first
splitting $H$ into a transverse part and a part along the $Z$-direction. The 
transverse
Hamiltonian 
\be
H_T=-\frac{1}{2m^*}[\frac{1}{R}\frac{\pt}{\pt R}( R 
\frac{\pt}{\pt R} ) +\frac{1}{R^2}{\cal D}_\Th^2]
 - \frac{1}{2m^*}\bbox{\sigma\cdot B}
\label{ht}
\ee
is still expressible as a square of a Dirac operator,
\bea
H_T &=& \frac{1}{2m^*}\left( \begin{array}{cc} \tilde{D}\tilde{D}^\dagger &0\\
0 & \tilde{D}^\dagger \tilde{D}  \end{array} \right) 
=\frac{1}{2m^*} H^2_{\tilde {D}},\nonumber\\
H_{\tilde {D}} &=& \left( \begin{array}{cc} 0 & \tilde{D} \\
\tilde{D}^\dagger  & 0 \end{array} \right)
\label{def2}
\eea
where
\bea
\tilde{D}= ie^{-i\Th} [  \frac{\pt}{\pt R}  -\frac{i}{R}
{\cal D}_{\Th}],\nonumber \\ 
\tilde{D}^{\dagger} = i e^{i \Th} [ \frac{\pt}{\pt R} +
\frac{i}{R}
{\cal D}_{\Th}].
\label{def3}
\eea

Requiring that $\tilde{D}$ and $\tilde{D}^\dagger$ be the 
adjoint of each other 
( so that $H_{\tilde{D}}$ is self-adjoint) leads once again to the 
same boundary conditions (\ref{2.8}). 

We will make the ansatz that the wavefunction is of the form
\be
\Psi = \psi_T e^{ikZ}.
\label{deaf}
\ee

   Now, the zero energy modes of $H_{\tilde{D}}$ ( and hence those of $H_T$) 
[ignoring the boundary condition at the origin and square integrability 
for the moment] are given (up to an overall 
normalization factor) by 
\be
\psi_T = \left( \begin{array}{c} C R^{m}e^{-\int^R_0 A_\Th (r') dr'}e^{i(m + 
\frac{\bt}{\al} k)\Th} \\
D R^{-(m +1)}e^{\int^R_0 A_\Th (r') dr'}e^{i(m + \frac{\bt}{\al} k+ 1)\Th}
 \end{array} \right).
\label{def4}
\ee

The corresponding functions for the full Hamiltonian (\ref{def1})
are plane waves along the $Z$-direction and are
given by 
\be
\Psi = \left( \begin{array}{c}C  R^{m}e^{-\int^R_0 A_\Th (r') dr'}e^{i(m +
\frac{\bt}{\al} k )\Th}
 e^{ik Z}\\D R^{-(m+1) }
e^{\int^R_0 A_\Th (r') dr'}e^{i(m +\frac{\bt}{\al} k +1 )\Th}
 e^{ik Z} \end{array} \right ) 
\equiv \left( \begin{array}{c}C R^{(m' - \frac{\bt}{\al} k)}
e^{-\int^R_0 A_\Th (r') dr'}e^{ i m' \Th}
 e^{ik Z}\\D R^{-(m' - \frac{\bt}{\al} k +1)}e^{\int^R_0 A_\Th (r') dr'}e^{ i
( m' +1)\Th}
 e^{ik Z}  \end{array} \right )
\label{def4a}
\ee
where $m' = m+\frac{\bt}{\al} k$.

The case where both $C$ and $D$ are nonzero cannot occur as we 
shall see below.

The quasi-periodicity condition (\ref{cone5})
leads to the following ``quantization'' conditions: 
\bea
&&\mbox{i) When $D=0$}: \qquad  \al  m'= \frac{\lambda}{2\pi} + n , 
\qquad n \in Z \!\!\! Z, \nonumber \\
&&\mbox{ii) When $C=0$}: \qquad  \al ( m'+1) = \frac{\lambda}{2\pi} + n , 
 \qquad n \in Z\!\!\!Z .
\label{def5}
\eea

For positive $\Phi$ , the requirement of  
square integrability at the 
origin $( r \ra 0)$ and at infinity $( r \ra \infty)$
and satisfying the boundary condition (\ref{2.9}) leads to the wavefunctions 
\be
\Psi (R, \Theta,Z)  = 
\left( \begin{array}{c}C R^{m} 
e^{- \int_0^R A_\Th  (r')dr'}e^{i(m + \frac{\bt}{\al} k )\Th } e^{ik Z}\\
0 \end{array}\right),
\label{def5.5}
\ee
as well as to bounds analogous to  (\ref{2.13}) : 
\bea
\mbox{a) When } e^{iK} \neq 1  :\qquad -\frac{1}{2} &<&  (m'-\frac{\bt}{\al}
 k) < 
( \frac{\Phi}{2\pi} - 1) ), \nonumber \\
\mbox{b) When } e^{iK} = 1 : \qquad -\frac{1}{2}&\leq& (m'- \frac{\bt}{\al} k)  < ( \frac{\Phi}{2\pi} - 1 ).
\label{def6}
\eea

When $\Phi$ is negative, the above requirements lead
to the wavefunction 
\be
\Psi (R, \Theta,Z)  = 
\left( \begin{array}{c}0 \\D R^{-(m+1)} 
e^{ \int_0^R A_\Th(r')dr'}e^{i(m + \frac{\bt}{\al} k +1)\Th } e^{ik Z}
 \end{array}\right),
\label{def6.5}
\ee
instead, and the bounds are now the analogs of (\ref{2.13a}):
\bea
\mbox{c) When } e^{iK} \neq -1: \qquad -\frac{1}{2} &>&  (m'-\frac{\bt}{\al} k) > 
 \frac{\Phi}{2\pi} , \nonumber \\
\mbox{d) When } e^{iK} = -1 :\qquad -\frac{1}{2} &\geq& (m'- \frac{\bt}{\al} 
k)  > 
\frac{\Phi}{2\pi}.
\label{def7.5}
\eea

The energy eigenvalues associated with these wavefunctions are
\be
E(m, k)= \frac{k^2}{2m^*}.
\ee 

The electronic states associated with the above 
wavefunctions 
will be loosely called zero mode electrons. The energy spectrum of these
zero modes is given by 
\be
E=\frac{k^2}{2 m^*}= \frac{\al^2 (m'-m)^2}{2m^* \bt^2} = 
\frac{(\frac{\lambda}{2\pi} + n - \al \bar{m})^2}{2m^* \bt^2}, 
\label{def10}
\ee
where 
\be
\bar{m} = \left\{ \begin{array}{l} m \qquad \mbox{for
$\Phi > 0$ ($D=0$)}, \\
m+1  \mbox{ for   $\Phi < 0$ ($C=0$)}.
\end{array} \right.
\ee
There are degenerate levels which can be obtained by varying the integer 
$n$ or alternatively by varying $m'$ while holding $n-\al m$ and hence $m'-m$ 
fixed.

In order to obtain the low energy
 effective action for the modes localized near the defect,
we need to look at the behavior of the solutions near the defect.
For  $\Phi > \pi$ for example, the probability densities for 
the wave functions with 
$m = -\frac{1}{2}$  are peaked around the origin
and  for those with  $m > -\frac{1}{2} $ vanish at the origin. A similar 
situation prevails for $\Phi < -\pi$.
These peaked states are allowed only if $e^{iK} = \pm 1$.
 Apart from these particular states , the other states are localized 
at a finite distance away from the defect line. However, in a realistic
situation , the defect line has a finite width and this will make some 
more  states to be localized within the defect line, provided 
$m$ is not too large and the magnetic field is 
large and can be treated as being uniform over the 
defect. This can be established from the Larmor formula,
\be
r^* = \frac{m}{B}
\label{others}
\ee
where $r^*$ is the radius of the Larmor orbit. The number of these 
low-energy excitations will be finite.

Though the above picture gives the single particle spectrum, 
one has to remember 
that in the many body picture for the fermions, 
one has to rather deal with the low energy 
excitations above Fermi surfaces.  They can be approximated by 
linearizing the theory around each Fermi surface.

Now we would like to see whether these excitations are chiral as a 
consequence of the chirality of the screw defect. In other words, we want to 
know if there can be some
asymmetry between excitations ``above''  the two Fermi surfaces
given by
\bea
k_F = \pm \sqrt{2 m^* E_F} \equiv \pm |k_F|, \nonumber \\
E_F = \frac{k_F^2}{2m^*}.
\label{tfermi}
\eea

The quasiparticle excitations are indeed chiral, as we can see in the 
following manner: 
We define their momenta as $\kappa \equiv k - 
k_F$. For excitations ``above''  the Fermi surfaces ( which we will 
call particles ) one has, 
\be
\Delta E = E - E_F \approx \frac{k_F}{m^*} \kappa > 0.
\label{excite}
\ee
Thus for particle excitations  at $k_F= |k_F|$, one has 
$\kappa > 0$  ( upward or ``right-movers'' ) and for those ``above''
the Fermi surface at $k_F = - |k_F|$, one has $\kappa < 0$ ( downward  
or ``left movers'' ).

 However, though the quasiparticles themselves are chiral, if the
situation is such that the number of upward moving and downward moving 
quasiparticles are equal, there will not be any net chirality. Let us investigate 
this point further. 

{\it For specificity, we 
assume that $\Phi > \pi$ and $\frac{\bt}{\al} >0$ 
in  the rest of the section}. 
Similar considerations
can be made for the remaining cases unless $\frac{\bt}{\al}=0$. We 
will  comment on the latter case later. 

\begin{figure}
\epsfxsize=6in
\epsfysize=8in
\epsfbox{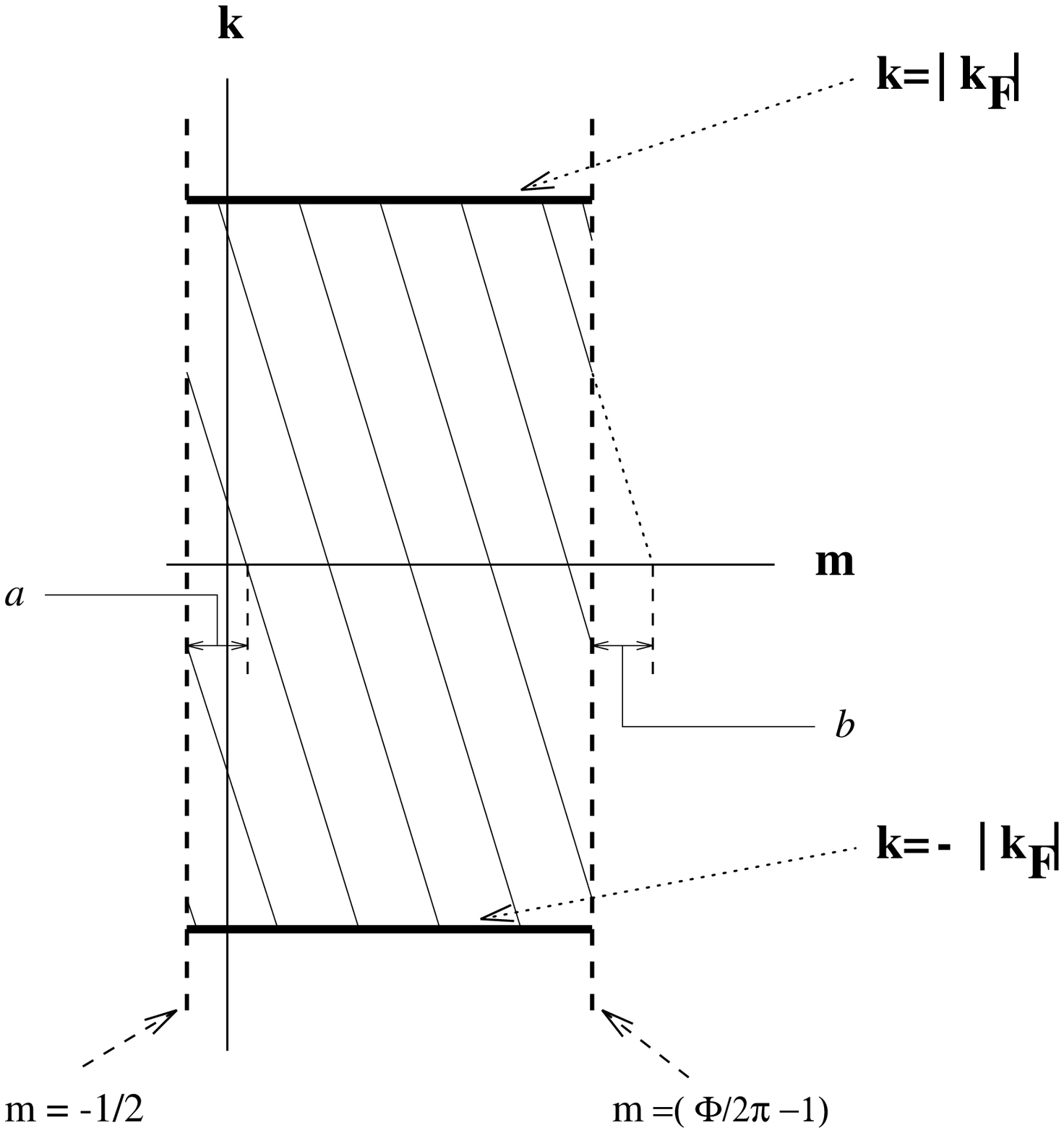}
\caption{ Plot of allowed $k$ vs. $m$ ( $\Phi > \pi$ ). Different slanted
lines correspond to different values of $m'$. }
\label{figure3}
\end{figure}

The relation between the linear momentum  along the $z$-axis 
and the angular momentum 
is given by
\be
m = m' - \frac{\bt }{\al} k
\label{angu}
\ee
 where $m $ has to satisfy the bound (\ref{def6}). 
 Also we have the fact 
that 
\be
\al m ' = \frac{\lambda}{2\pi} + n , \qquad n  \in Z \!\!\! Z.
\label{quant}
\ee
This shows that for a fixed value of $k$, the angular momenta
of two neighboring states differ by 
\be
\Delta m' = \frac{1}{\alpha}.
\label{differ}
\ee
In  Figure \ref{figure3} we have plotted the relation between 
 $k$ and  $m$ for different $m'$s. 

The lowest energy electrons occupy the levels lying {\it  within}  the   
region bounded by the lines $m = -\frac{1}{2}$ and $m = \frac{\Phi}{2\pi} 
-1$ and $k = \pm |k_F|$, which we will call the Fermi sea hereafter.

We would like to determine the number of straight lines ( which are 
labeled by different values of $m'$ )  crossing  
the two Fermi surfaces. These lines correspond to the 
low energy excitations of the systems above the Fermi surfaces.

Note that the relation (\ref{angu}) shows that as we increase $k$ ( 
by applying external electric field, for instance ) , 
$m$ decreases for fixed $m'$. As 
we have argued 
earlier that $m$ is a measure of how far away from the defect the 
electron is localized , this shows that the electron is drawn into the 
defect as $k$ is increased. ( Note that the direction of this  current 
is dictated by the sign of $\frac{\bt}{\al}$ and
that this transport phenomenon will not  occur 
if $\frac{\bt}{\al} =0$ ).

Using relations (\ref{angu}) and (\ref{differ}), one can now easily 
estimate the number of occupied states with a given momentum
[the number of intersections of the 
graphs in Figure \ref{figure3} with a horizontal line with a given 
$k$], provided 
we know the maximum and minimum value of $m$ for the 
states lying 
within our allowed region  when $k = 0$, which we denote 
by $m_{max}$ and $m_{min}$ 
respectively. Thus, the number of excitation branches when $k= 0$ is
\be 
N_0 \equiv  \al ( m_{max} - m_{min}) + 1.
\label{number}
\ee

Let us now define the following quantities

\bea
a &=& m_{min} + \frac{1}{2}, \nonumber \\
b&=& m_{max} + 1 - (\frac{ \Phi}{2\pi} -1 ) 
= ( m_{max} +2 - \frac{\Phi}{2\pi}).
\label{ab}
\eea

The quantity $a$ measures the horizontal distance between the line $m= -
\frac{1}{2}$ and the occupied state which is just to the right of it for 
$k=0$, while 
$b$ measures the horizontal distance between the line 
$m= (\frac{\Phi}{2\pi} -1)$ and the state which is just
to the right of it for $k=0$.  (It moves into the Fermi 
sea  when 
$k$ is sufficiently increased).  The value of $a$ depends on the 
parameters $\lambda$ and $\al$ ands hence is fixed for a given 
material ( assuming that $\lambda$ cannot be varied ). On the other hand, 
$b$ depends on the flux $\Phi$ and hence can be varied. Neither $a$ nor $b$
are functions of $k_F$.
 
Now, as we increase $k$ from zero to a positive value, 
the $m$ values 
for the occupied states will change and those with $m \leq -\frac{1}{2}$ 
will ``flow'' out of the spectrum. A simple counting argument shows that 
the number of states that will ``flow'' out will be given 
by the integer
\be
N_1 = \left[ \frac{ k- \frac{\al a}{\bt}}{\frac{1}{\bt}} \right] 
= [ \bt k - \al a].
\label{out1}
\ee
 This equation can be obtained by noticing that the vertical spacing 
between two successive branches is $\Delta k = \frac{1}{\bt}$ and 
that the
minimum value of $k$ required to move the left-most allowed 
state to the left of the bound $m = -\frac{1}{2}$ is $\frac{\al a}{\bt}$.

At the other extreme of the allowed region, there will states moving ``into'' 
the allowed bound. The number of these incoming states can be found by 
arguments similar to those above. This number is given by
\be
N_2 = [ \bt k - \al b].
\label{in1}
\ee   

Thus the number of branches appearing through the top Fermi surface 
at $ k = |k_F|$ is given by
\be
N_{top} = N_0 - N_1 + N_2 = \al ( m_{max} - m_{min} ) +1 
-[ \bt |k_F| - \al a] + [ \bt |k_F| - \al b].
\label{top}
\ee

\begin{figure}
\epsfxsize=6.5in
\epsfysize=6.5in
\epsfbox{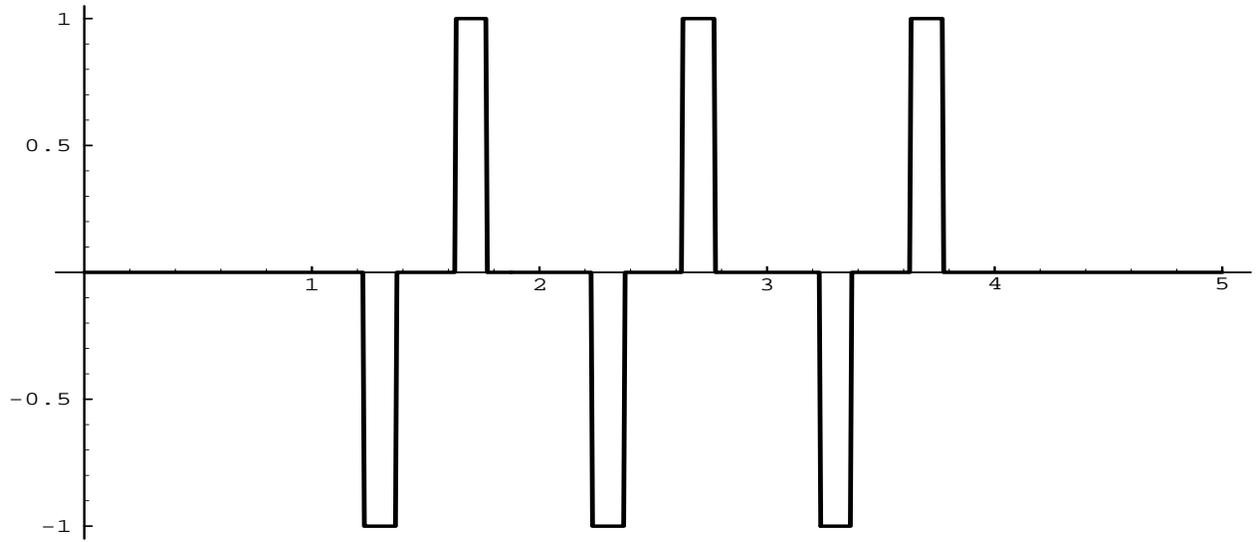}
\caption{ Plot of $\eta$ vs. $ \bt |k_F|$  for  $\al a~=~.23$ 
and $\al b~=~ .37$.}
\label{figure4}
\end{figure}

\begin{figure}
\epsfxsize=6.5in
\epsfysize=6.5in
\epsfbox{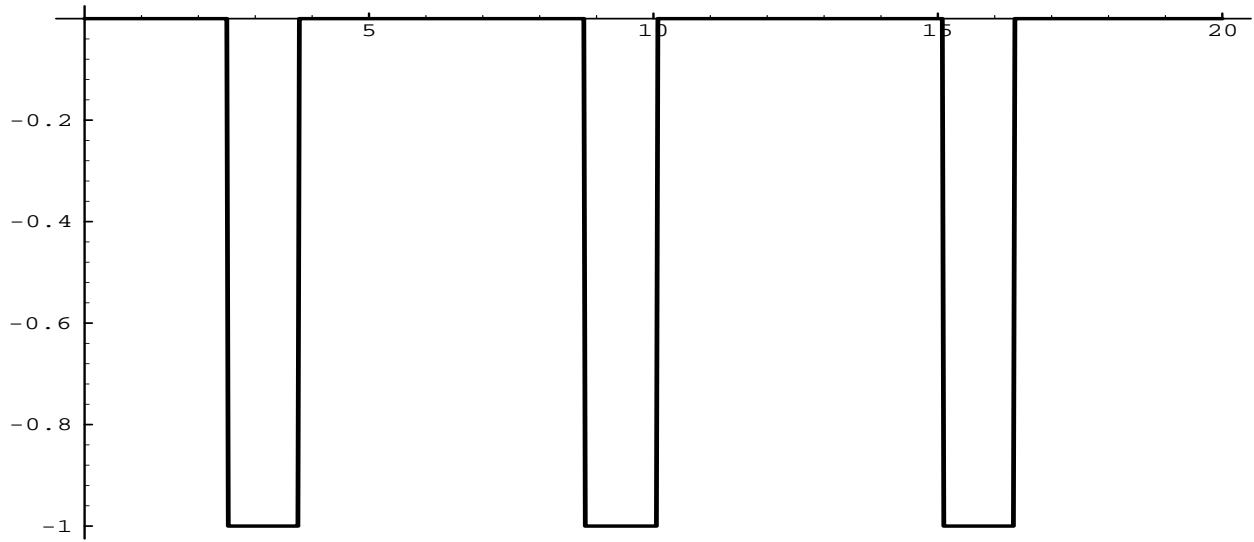}
\caption{ Plot of $\eta$ vs. $\Phi$  with $\al a=~.37$ 
and $\bt |k_F| = 5.7$ and $m_{max} = 5$.}
\label{figure5}
\end{figure}

We can also evaluate the number of branches appearing through the bottom 
Fermi surface at $k = - |k_F|$, employing procedure similar to above.
However, this time we have to decrease the $k$ value from zero 
 to a negative 
value $ k < 0$. Decreasing the $k$ value will make the $m$ value for 
various states to increase due to the relation (\ref{angu}) since 
  $\frac{\bt}{\al}$ is positive by  
assumption. Consequently, 
states will move away from the defect and some states 
with $ m \leq - \frac{1}{2}$ will
move into  the Fermi sea while some other states close to the bound 
$ m = \frac{\Phi}{2 \pi} -1$ will `` flow'' out of the Fermi sea. 
The number of branches  appearing out of the Fermi surface $k= -|k_F|$ is then 
given by 
\be
N_{bottom} = \al ( m_{max} - m_{min} ) + 1 + [ \bt |k_F| + \al a] - [ \bt 
|k_F|+ \al b].
\label{bottom}
\ee
Thus the difference between the number of states from the 
top and bottom Fermi surfaces is 
\be
\eta = N_{top} - N_{bottom} = [ \bt |k_F| - \al b] + [  \bt |k_F| + \al b]
-( [ \bt |k_F| - \al a]+ [ \bt |k_F| + \al a]).
\label{asymmetry}
\ee 
This number $\eta$ determines the total chirality of the excitations.
Note that when $a$ = $b$ ,  $N_0 = N_{top}= N_{bottom}$
 and $\eta = 0$. Thus, in this case the number of species remains the same irrespective 
of the value of $|k_F|$. However, if $a \neq b$ ,  $\eta$  
can take any one of the three values , +1, 0 and -1. 
This is shown in Figure \ref{figure4} where $\eta$ is plotted as a
function of $|k_F|$ for fixed values of $ \al a$ and $\al b$ ( which 
we have chosen arbitrarily).  $\eta$ can 
also change as the value of  $\al b$  is changed ( by varying $\Phi$ ) while  
$|k_F|$ and $\al a$ are kept fixed.
This is shown in Figure \ref{figure5}. So, the direction of the 
current due to the quasiparticles 
will depend on the values of $|k_F|$, $\al b$ and $\al a$. It depends 
in particular on $\Phi$.

We finally comment on the possibility $\frac{\bt}{\al}=0$. Since, $|\al|
 \leq 1$, we then have $\bt =0$ which implies the absence of the screw defect.
In this case, the straight lines in Figure \ref{figure3} are vertical and 
the net chirality $\eta$  is always zero.

The zero modes dominate the contribution to the effective action 
for the modes localized near the defect. We will analyze the low  
energy effective action describing these chiral mode  further in Section V.

\section{Transport Along Solenoidal Defects and Anomalies}

We have considered a situation where there are 
localized states
in the presence of a solenoid with flux $\Phi$.
We will now consider the response of these trapped states to a constant 
electric field along the axis of the solenoid ( that is,  the defect). 
The low lying excitations about the Fermi surface corresponding to these 
states are described by massless chiral fermions 
localized on the defect. We proceed to discuss
how their coupling to electromagnetism leads to 
anomalous conduction properties. To set up the relevant background for this
discussion, we first recall the reasons for the $U(1)$ anomaly  of  chiral \
$(1+1)$ dimensional gauged fermionic systems.

    Consider  $(1+1)$ dimensional chiral , say right-handed fermions, 
with field, $\psi_R = \frac{1}{2} ( 1+ \gamma_5)\psi$. The action of this  
 fermion field coupled to a constant  electric field $E ( > 0)$ is given 
by 
\be
S = \int dt dx {\bar{\psi}}_R \gamma^\mu ( i\pt_\mu + A_\mu) \psi_R.
\label{an1}
\ee
We choose the following
$\gamma$ matrix convention:
\be
\gamma^t= \sigma_1; \quad ~ \gamma^x= -i \sigma_2,
\label{gamma2}
\ee
so that $\gamma_5 =  \gamma^t \gamma^x = \sigma_3$. The Dirac Hamiltonian 
for the 1+1 dimensional massless fermion is 
\be
\tilde{H} = \sigma_3  (p- A_1) - A_0.
\label{dirham}
\ee
Heisenberg's equation of motion gives, for the velocity $v$,
\be
\dot{v} = \frac{d}{dt}(p- A_1) = i [ \tilde{H}, p - A_1] + \frac{\partial}{\partial\, t}(p-
A_1)  = - E .
\label{fer}
\ee
 
The derivation of the chiral $U(1)$ anomaly for the 
action (\ref{an1}) by Nielsen and Ninomiya \cite{NN} is as follows. 
The density of states $\rho (v)$ for a massless $(1+1)$ dimensional
particle is controlled by the following formula for its variation :
\be
\delta  \rho(v)= \frac{1}{2\pi} \delta v.
\label{an4}
\ee
Hence the production rate of these excitations per unit length 
is 
\be
\dot{N}_R(t) = \frac{d \rho}{ dt } = \frac{1}{2\pi} \dot{v} = \frac{ E}{2 \pi}
\label{an5}
\ee
where $N_R(t)$ is the
number of fermions per unit length and 
we have used (\ref{fer}).
As these excitations carry charge,
 there is a non-conservation of electric
charge $Q_R(t)$ carried by the right-handed fermions  
in the (1+1) dimensional theory, the rate of change 
of $Q_R(t)$ being  
\be
\dot{Q}_R(t)= e \dot{N}_R(t) L = -\frac{ E }{ 2 \pi}L
\label{an7}
\ee
where $L$ is the size of the sample.
In this way, we obtain the $U(1)$ anomaly. 

One can show in a similar way that for a left-handed chiral fermion, 
the rate of change of its charge $Q_L(t)$ is given by
\be
\dot{Q}_L(t)= e \dot{N}_L(t) L =  \frac{ E }{ 2 \pi}L
\label{an7.5}
\ee
where $N_L (t)$ is the number density of the left-handed fermions.

Thus if there are $n_1$ species of right-handed fermions 
and $n_2$ species of left-handed fermions, then the rate of change 
of  electric charge is  given by
\be
\frac{d}{dt} ( Q_R (t) + Q_L (t) ) \equiv \dot{Q}_{total} = -(n_1 - n_2 ) \frac{E}{2\pi} L.
\label{totanomaly}
\ee

As we are going to treat the different branches of the quasiparticle 
excitations as different species for the screw defect situation, one 
can readily see that 
\be
(n_1 - n_2 ) = \eta 
\label{difference}
\ee
where $\eta$ is given by (\ref{asymmetry}).

But one knows that the system as a whole has charge conservation. Therefore, 
when $\eta \neq 0$,  it 
must 
be that charge from outside flows into or away from the 
defect, and there must be 
non-conservation of charge outside the defect too. 
This fact can be encoded in an 
anomalous  Chern-Simons ``effective'' action, representing the electronic 
degrees of freedom that are not localized on the defect. 
The connection between the anomaly
(\ref{an7}) and the Chern-Simons term
 is elegantly  demonstrated in \cite{Harvey}( See
also \cite{balu}). 

\begin{figure}
\epsfxsize=6in
\epsfysize=4.5in
\epsfbox{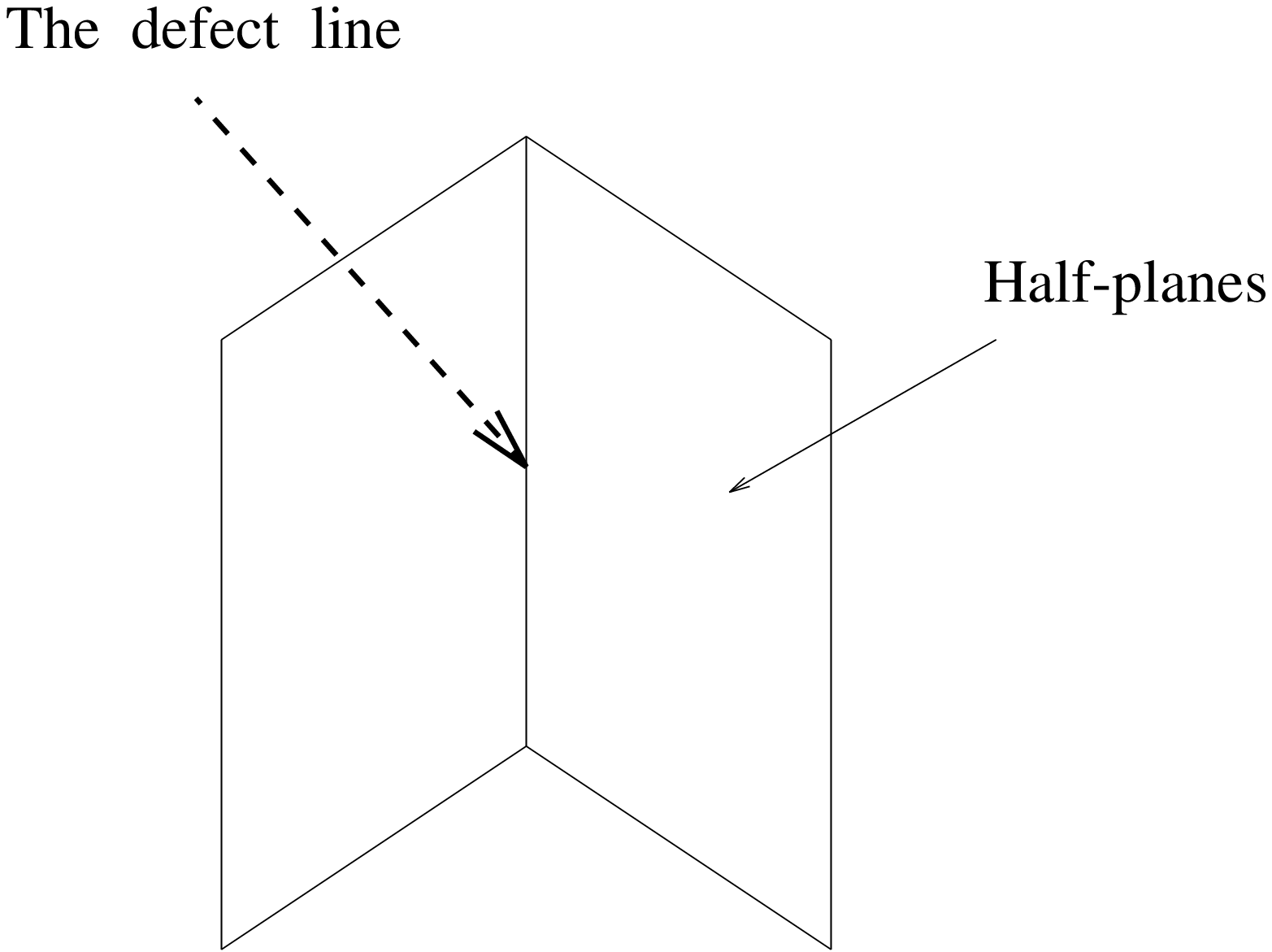}
\caption{  Half-planes with the defect as the boundary.}
\label{figure2}
\end{figure}

 Using their results, it can be shown that 
the contribution appearing on the right-hand side of the equation (\ref{an7})
is  cancelled by the bulk action given by
\be 
S_{bulk}= - \frac{\eta}{8 \pi^2} \int  d \phi \left[ \int dt ~r dr ~dz A dA 
\right],
\label{an12}
\ee
where $\eta$ is given by the equation (\ref{asymmetry}).
The Chern-Simons three-form 
is being  integrated here on an infinite half-plane which has 
the defect as the 
boundary, as shown in Figure \ref{figure2}. 
The gauge variation of the Chern-Simons term produces a surface
term which is exactly cancelled by the variation of 
the chiral anomaly on the defect  \cite{Harvey,balu},
the coefficient $\frac{1}{8\pi^2}$ in (\ref{an12}) is chosen that 
this cancellation occurs.

The fact that the Chern-Simons coefficient is determined by the 
anomaly fixes of the transverse conductivity 
of the defect, there being an exchange 
of current with the bulk when an
electric field is applied along the line defect. This transverse 
conductivity is given by 
$\frac{\eta}{4\pi^2} \times 2\pi = \frac{\eta}{2\pi}$, the $2\pi$ coming
from the $\phi$ integration.[ The first factor comes from 
varying $A$ \cite{balu}.  Recall that we have set the electric charge 
$e=1$, if we had not done so, this conductivity would have
been  $\frac{\eta e^2}{2\pi}$]. As the value of $\eta$ can change 
by $\pm 1$ only, the system will behave somewhat like 
the integer Quantum Hall samples thought here the conductivity is 
bounded and  the sign of the current is fixed by $\eta$ .

\section{Conclusion}
In this paper we have studied the effect of line defects on the transport 
properties of solids in the presence of a magnetic 
flux parallel to the line. 
This was done by describing the defects in a crystal 
in terms of a non-trivial metric and then studying the propagation of 
electrons in this background metric. We have also 
discussed the low energy effective
action for this system. It is a $1+1$ dimensional chiral fermionic action 
 localized 
on the defect. The coupling of this system to an external electric field 
parallel to the defect leads to 
an anomaly and induces a bulk action involving the Chern-Simons term.
The resultant total action  
in turn leads a phenomenon 
similar to the transverse 
Hall effect with a quantized conductivity. However, the sign of the 
conductivity, which is proportional to $\eta$, undergoes flips as one changes the value of the magnetic flux 
or the Fermi momentum .

There are numerous interesting ideas suggested by the description of 
defects in terms of metrics and we 
will now discuss a few of them. It has been shown that in the 
background of disclination defects, electric 
charges experience a non-trivial force which arises due to the conical 
defect angle and is directed towards, or away from, the 
defect line. For a positive conical angle ( where a wedge has been 
taken out ), this force turns out to be attractive ( that is, towards the 
defect line )  and  
leads to bound states of electrons and defects \cite{fernando}.
It would be interesting to see how this force is
modified by the screw dislocation. 

We have assumed here
that the magnetic field is  localized in the neighborhood of the defect.
In a type-II superconductor such a scenario can actually 
happen. The strain energy due to the defect will force the 
region around the defect to be in the normal state. Then 
 it is favorable  energetically for the magnetic flux tubes 
to be trapped within the 
screw defects ( if they are present in the material) \cite{cristina}. 
This is basically due 
to Meissner effect. Our results, therefore,  would be interesting in the 
study of flux-line trapping in superconductors by defects\cite{cristina}.

In this paper we have discussed the fact that  electronic
conductivity transverse to the defect
is fixed in quantized units because the low energy effective theory of the
electrons localized on the defects has an 
electromagnetic anomaly. 
It is known however that there is a gravitational anomaly in the same
(1+1) dimensional system. 
It will lead to a transverse ``gravitational Hall effect'', and
this will manifest itself in anomalous elastic and vibrational
properties. Work is in progress to analyze in detail how this affects
the properties of the system. 

The fact that static defects can be described in terms of stationary
solutions to 2+1 dimensional Einstein action for gravity 
with matter gives rise to the
interesting possibility that the dynamics of these defects can be
modeled in terms of a fully dynamical theory of gravity.
It would be interesting to carry out the
study of dynamics of defects in the language of quantum gravity
in two and three dimensions.  In particular it suggests 
 the interesting possibility that systems with defects
 could provide us with analog simulations of many situations in quantum 
gravity that 
are now being studied using elaborate computer simulations.

The interplay of ideas from two widely
different fields like condensed matter physics and quantum gravity
can lead to lots of new results of the kind we have just
discussed. It would be very interesting to use these condensed matter
systems as  low energy experimental probes into what is
conventionally regarded as the domain of Planck scale physics. Similar analog 
probes of cosmological defects have already led to rewarding results 
(\cite{schiff})

\section*{Acknowledgment}

We would like to thank E. Dambasuren, P. Golumbeanu,
G. Jungman,
S. Vaidya for furious discussions in  Room 316. A.M. would also like to 
thank M. C. Marchetti for discussions and for pointing out  reference
\cite{cristina}. 
This work was supported by the US Department of Energy under contract 
number DE-FG02-85ER40231.

\newpage
\newpage
\newpage

\end{document}